\documentstyle[11pt,aaspp4,flushrt]{article}

\newcommand{\mjyb}{mJy~beam$^{-1}$}
\newcommand{\mum}{$\mu$m}
\newcommand{\cm}{cm$^{-3}$}
\newcommand{\hi}{H{\sc I}}
\newcommand{\hii}{H{\sc II}}
\newcommand{\uchii}{UC~H{\sc II}}

\newcommand{\Nc}{$N'_c$}

\newcommand{\lb}{($l$, $b$)}
\newcommand{\mins}{$\!'$}

\newcommand{\kms}{km~s$^{-1}$}
\newcommand{\vlsr}{$v_{\rm LSR}$}
\newcommand\co{$^{12}$CO}
\newcommand\coj{$^{12}$CO J=1$-$0}
\newcommand\coo{$^{13}$CO}
\newcommand\cooj{$^{13}$CO J=1$-$0}
\newcommand\ratioa{$^{12}R_{2-1/1-0}$}
\newcommand\ratiob{$^{12/13}R_{1-0}$}
\newcommand\css{C$^{34}$S}
\newcommand\csj{CS J=2$-$1}

\newcommand\Ncoo{$N(^{13}{\rm CO})$}
\newcommand\Ncs{$N({\rm CS})$}
\newcommand{\Nhtwo}{$N({\rm H_2})$}
\newcommand{\nhtwo}{$n({\rm H_2})$}

\newcommand{\mlte}{$M_{\rm LTE}$}

\newcommand{\lir}{$L_{\rm IR}$}
\newcommand{\msol}{$M_\odot$}
\newcommand{\lsol}{$L_\odot$}
\newcommand{\ta}{$T_{\rm A}^*$}
\newcommand{\tb}{$T_{\rm b}$}
\newcommand{\tRs}{$T_{\rm R}^*$}
\newcommand{\tr}{$T_{\rm r}$}
\newcommand\tex{$T_{\rm ex}$}
\newcommand{\delv}{$\Delta v$}
\newcommand\hrrl{H76$\alpha$}

\begin{document}


\title{Interaction between Ionized and Molecular Gas\\
in the Active Star-Forming Region W31}

\author{Kee-Tae Kim and Bon-Chul Koo}

\affil{Astronomy Program, SEES, Seoul National University, Seoul
151-742, Korea:\\
kimkt@astro.snu.ac.kr, koo@astrohi.snu.ac.kr}



\begin{abstract}

We have carried out 21~cm radio continuum, \hrrl\ radio recombination line,
and various (\co, \coo, CS, \& \css) molecular line observations of the W31 
complex. 
Our radio continuum data show that W31 is composed of two extended
\hii\ regions, G10.2$-$0.3 and G10.3$-$0.1, each of which comprises 
an ultracompact \hii\ region, two or more compact components, 
and diffuse envelope.
The W31 cloud appears as an incomplete shell on the whole
and consists of southern spherical and northern flat components,
which are associated with G10.2$-$0.3 and G10.3$-$0.1, respectively.
For an assumed distance of 6~kpc,
the molecular cloud has a size of 48~pc and a mass of 6.2$\times$10$^5$~\msol.
The IR luminosity-to-mass ratio and the star formation
efficiency are derived to be 9~\lsol/\msol\ and 3\%,
respectively. These estimates are greater than average values 
of the inner Galactic plane. 
We detect two large (16 and 11~pc) and massive (2.1$\times$10$^5$ and
8.2$\times$10$^4$ \msol) CS-emitting regions in the northern and
southern cloud components.
The large amount (48\% in mass and 16\% in area) of dense gas may suggest 
that the W31 cloud has ability to
form rich stellar clusters and that star formation has only recently begun.
The extended envelopes of both G10.2$-$0.3 and G10.3$-$0.1 are likely to
be results of the champagne flows,
based on the distributions of ionized and molecular gas and
the velocity gradient of \hrrl\ line emission.
According to the champagne model, the dynamical ages of the two \hii\
regions would be (4$-$12)$\times$10$^5$~yr.
We find strong evidence of bipolar molecular outflows associated with
the two ultracompact \hii\ regions.
In the vicinity of the ultracompact and compact \hii\ regions in G10.3$-$0.1,
the \co\ J=2$-$1/J=1$-$0 intensity ratio is high (1.4) 
and a small but prominent molecular gas hollow exists. 
Together these observations strongly indicate that 
the \hii\ regions and their ionizing stars are interacting with
the molecular cloud.
Therefore, it is most likely 
that recently formed massive stars are actively disrupting 
their parental molecular cloud in the W31 complex.

\end{abstract}


\clearpage
\section{Introduction}

Stars form in the dense cores of molecular clouds 
and eventually emerge as visible objects. 
Thus the molecular gas and dust surrounding newly formed stars should be 
disrupted.
Massive stars are expected to play an important role
in dispersing the ambient materials,
since they emit strong radiation, stellar winds, and outflows
even at their earliest evolutionary stage.
Moreover, massive stars could dramatically impact on 
the surrounding environments 
through \hii\ regions and supernova explosions.
However, direct observational evidence of this conjecture
is rarely seen.

W31 is one of the brightest \hii\ region/molecular cloud complex
in the inner Galaxy.
Kalberla, Goss, \& Wilson (1982) suggested 
on the basis of their sensitive \hi\ absorption line observations 
that W31 is associated with the expanding 3-kpc arm with a peculiar
velocity of $\sim$30~\kms\ and so lies 
about 6~kpc from the Sun (see also Wilson 1974). 
We adopt this distance throughout the present paper,
although there are some reports that propose different distances,
e.g., 14.5$\pm$1.4~kpc (Corbel et al. 1997) 
and 3.4$\pm$0.3~kpc (Blum, Damineli, \& Conti 2001). 
W31 appears as two prominent extended \hii\ regions, 
G10.2$-$0.3 and G10.3$-$0.1, 
in the radio continuum maps with low ($>$1$'$) angular resolutions
(e.g., Reich, Reich, \& F\"{u}rst 1990).
High-resolution ($<$10$''$) radio continuum observations show that
the central regions of both \hii\ regions form elongated 
ionization ridges (Woodward, Helfer, \& Pipher 1984; Ghosh et al. 1989).
Two ultracompact (UC) \hii\ regions, G10.15$-$0.34 and G10.30$-$0.15, 
were found to be located at the peaks of the ionization ridges, respectively
(Wood \& Churchwell 1989; Kim \& Koo 2001).
Several other massive young stellar objects were also identified 
in G10.2$-$0.3 by near-infrared observations (Blum et al. 2001).
These suggest that massive star formation
is actively ongoing in the W31 complex.
On the other hand,
a Crab-like supernova remnant (SNR) G10.0$-$0.3 is $\sim$10$'$ southwest of
G10.15$-$0.34. 
The SNR  is likely to be associated with a soft $\gamma$-ray repeater,
SGR 1806$-$20 (Kulkarni et al. 1994).
Accordingly, the W31 complex seems to be one of the best laboratories 
for studying the formation of massive stars and
the interaction of massive stars with their parental molecular cloud.

In this paper,
we present radio continuum, radio recombination line (RRL), and 
various molecular line observations of the W31 complex.
Our observations aim 
to investigate the physical characteristics of molecular clouds
that are actively forming massive stars, and
to explore how massive stars and their natal molecular clouds
interact with each other.
The observations are described in \S~2 and the results are presented in \S~3. 
In \S~4 we discuss the physical properties of the W31 molecular cloud 
and the interaction between ionized and molecular gas in the region. 
Our main results are summarized in the last section.

\section{Observations}

\subsection{Radio Continuum and Recombination Lines}

The radio continuum and RRL data are extracted
from the survey of 16 \uchii\ regions with extended envelopes.
The observational details were presented by Kim \& Koo (2001).
In summary, radio continuum observations were made at 21~cm with
the VLA of
the NRAO\footnote{
The National Radio Astronomy Observatory is
operated by Associated Universities, Inc., under cooperative
agreement with the National Science Foundation
} 
in 1995 February.
The array was in the DnC hybrid configuration and the observations
were sensitive to structures up to 15$'$.
The final map resulted from two separate pointings of the array. The phase
centers were set at G10.15$-$0.34 and G10.30$-$0.15, 
i.e., $(\alpha, \delta)_{1950}$=
($18^{\rm h} 06^{\rm m} 22.^{\!\!\rm s}5$, $-20^\circ 20' 05''$)
and
($18^{\rm h} 05^{\rm m} 57.^{\!\!\rm s}9$, $-20^\circ 06' 26''$)
or $(\alpha, \delta)_{2000}$=
($18^{\rm h} 09^{\rm m} 21.^{\!\!\rm s}0$, $-20^\circ 19' 31''$)
and
($18^{\rm h} 08^{\rm m} 56.^{\!\!\rm s}1$, $-20^\circ 05' 53''$).
The data were calibrated, imaged, and mosaicked using the AIPS.
The resulting synthesized half-power beamwidth was $37'' \times 25''$
and the rms noise level of final image was 1.7 mJy~beam$^{-1}$.

\hrrl\ (14.68999 GHz) RRL observations were conducted 
using the 140 foot telescope of the NRAO in 1997 February and June. 
The telescope has a FWHM of about 2$'$ and 
a main beam efficiency of 0.58 at the observing frequency.
Both circular polarizations were observed simultaneously using two 1024
channel autocorrelators with 40~MHz bandwidth each, yielding a velocity
resolution of 1.59~\kms\ after Hanning smoothing.

\subsection{Molecular Lines}

We have carried out \co, \coo, CS, and \css\ line observations of 
the W31 complex. A summary of the observations is given in Table 1.
\cooj\ and \co\ J=2$-$1 line observations were made with the 12~m 
telescope of the NRAO at Kitt Peak using the on-the-fly mapping technique. 
We used 256 channel filterbanks with 64~MHz and 128~MHz bandwidths
as the backend, respectively.
A squared area of $30' \times 40'$ centered at
$(\alpha, \delta)_{1950}$= 
($18^{\rm h} 06^{\rm m} 09.^{\!\!\rm s}1$, $-20^\circ 09' 23''$)
was mapped at  20$''$ spacing in the \cooj\ line,
while a $5' \times 5'$ region centered at G10.30$-$0.15 
was mapped at 10$''$ sampling in the \co\ J=2$-$1 line.
We used positions with some emission as reference positions,
\lb\ = ($10.\!\!^\circ15$, $-1.\!\!^\circ50$)
and ($10.\!\!^\circ05$, $+0.\!\!^\circ03$),
because we could not find any positions devoid of \co\ and/or \coo\ emission 
near the source.
We obtained a sensitive spectrum of each reference position
and then added it to the position-switched spectra.
The system temperatures were typically 300~K at 110~GHz and 400~K at
230~GHz.
We have converted the observed temperatures (\tRs) to the
main-beam brightness temperature (\tb) using the corrected main-beam
efficiencies ($\eta_{\rm mb}^*$) provided by the NRAO.

CS J=2$-$1 line observations were undertaken using 
the 14 m telescope of the Taeduk Radio Astronomy Observatory (TRAO).
We mapped two regions  of $11' \times 14'$ and $12' \times 10'$,
respectively, centered at G10.15$-$0.34 and G10.30$-$0.15
in full-beam (60$''$) spacing.
The two \uchii\ regions were observed in the CS J=3$-$2 and
\css\ J=2$-$1 and J=3$-$2 lines as well.
We used SIS mixer receivers with 256 channel filterbanks of 64~MHz bandwidth. 
The spectra were obtained by position switching to
\lb\ = ($10.\!\!^\circ15$, $0.\!\!^\circ50$),
which was checked to be free from appreciable ($T_{\rm A}^* < 0.1$~K)
CS emission.
The system temperature varied in the range 300$-$500~K at 98~GHz and 
600$-$800~K at 147~GHz.
We also made \coj\ line observations of the region mapped in \co\ J=2$-$1
line emission using a 256 channel filterbank with 256~MHz bandwidth.
The system temperature was about 1000 K.
The line intensity was obtained on the \ta\ scale from the chopper-wheel 
calibration.

\section{Results}

\subsection{Radio Continuum and Recombination Line Results}

Figure 1 is our 21~cm radio continuum image of the W31 complex.
There are two distinct extended \hii\ regions, G10.2$-$0.3 and G10.3$-$0.1,
in the field. 
G10.2$-$0.3 has several compact components, almost all of which
are embedded in low-level extended envelope of
$10.\!'9 \times 6.\!'7$ (or 19.0~pc $\times$ 11.1~pc).
The central region is elongated in the east-west direction on the whole
and is composed of two strong compact components. 
The \uchii\ region G10.15$-$0.34 lies at the peak of the western one.
High-resolution ($\sim$5$''$) VLA continuum observations at 5~GHz 
suggest that the central region comprises about 20 dense ionized clumps 
(Woodward et al. 1984).
The surface brightness declines sharply to the west, 
while it decreases slowly to the east. 
Figure 2 shows \hrrl\ spectra observed along
$\delta (1950) \simeq -20^\circ 20' 05''$.
The center velocity decreases with increasing right ascension.
The velocity difference is about 10~\kms\ between the central region
and the eastern protuberance. 
 
In G10.3$-$0.1, there are one isolated compact component
and two strong ones surrounded by diffuse emission extending over 
$12.\!'8 \times 4.\!'6$ (or 22.3~pc $\times$ 8.0~pc).
The two central compact components form an ionization ridge elongated 
in the northeast-southwest direction. The \uchii\ region G10.30$-$0.15
is located at the peak of the western one.
The diffuse envelope extends straight in the northwest-southeast direction,
which is perpendicular to the central ridge.
There is no significant velocity difference
between the central region and the envelope.
According to Kim \& Koo (2001),
the \uchii\ regions and their associated compact components are likely
to be excited by the same ionizing sources in both G10.2$-$0.3 and G10.3$-$0.1,
while the individual compact components may be ionized by separate
sources.

\subsection{\cooj\ Line Results}

Figure 3 exhibits \cooj\ line profiles observed towards G10.15$-$0.34
and G10.30$-$0.15.  There are three or more velocity components in the lines 
of sight. In each spectrum, the strongest component is associated
with the \uchii\ region. The center velocities of \coo\ gas are
shifted by 5$-$8~\kms\ from those of the ionized gas, 
marked by the vertical dotted lines.
There seem to be high-velocity wings in the \coo\ lines of both \uchii\
regions, even though the features are significantly confused by multiple 
velocity components. 
Shepherd \& Churchwell (1996) have also found high-velocity wings in 
the \coj\ line of G10.30$-$0.15, but could not identify them 
in G10.15$-$0.34 due to a complex blend of lines. 
We will discuss in more detail the high-velocity gas using our CS spectra 
in the next section.

Figure 4 is \coo\ line integrated intensity map, 
obtained by integrating over the velocity range between 
\vlsr\ = 0 and 22~\kms. 
Large crosses indicate the positions of \uchii\ regions, 
while small crosses represent the peak positions of the strong compact 
components without \uchii\ regions (see Fig. 1). 
The molecular cloud is very clumpy. There is a large hollow over 
the central region, which makes the cloud appear 
as an incomplete shell being open to the northeast. 
A possible explanation for this morphology is that
the cloud has been produced by energetic events near the center, 
such as strong stellar winds and/or supernova explosions of massive stars,
although we have not found any massive star or nonthermal
radio continuum emission in the hollow region.
As will be discussed in \S~4.2,
the dynamical age of G10.2$-$0.3 is similar to that of G10.3$-$0.1,
suggesting that star formation has begun almost simultaneously 
at the southern spherical and northern flat components.
This may be in favor of the above interpretation.
It seems that the northern component has been substantially dispersed by massive stars,
since a small but conspicuous hollow exists near the UC and compact
\hii\ regions,  $(\alpha, \delta)_{1950}$ $\approx$
($18^{\rm h} 05^{\rm m} 58^{\rm s}$, $-20^\circ 05' 45''$) (see \S~4.2). 
In the southern component, on the contrary, 
the dense \coo\ cores associated with the UC and compact \hii\ regions 
are still strong.
It is worthwhile to note that the integrated intensity drops steeply 
on the eastern boundary of the southern component,
in contrast to the radio continuum intensity distribution.

Figure 5 displays the channel maps of the molecular cloud.
\coo\ emission appears mainly in the velocity range 
between 2 and 20~\kms.
The channel maps show that the northern and southern components are 
certainly associated, 
and reveal more clearly incomplete shell-like morphology of the cloud.
The molecular gas shell remains nearly constant in size 
as the velocity varies, suggesting that it does not dynamically 
expand or contract. 
There are small velocity variations in both northern and southern
components. As the velocity decreases, strong \coo\ emission moves from
the western part to the eastern part in the northern component, 
while it moves from the northeastern part through the northern part 
to the southern part in the southern component.
The \coo\ core associated with G10.15$-$0.34 ($\sim$8~\kms) is slightly 
blueshifted from the main shell-like structure ($\sim$12~\kms).
The northern flat component seems to be nearly broken through at 
the position of the distinctive hollow, which persists over a velocity 
interval of 10$-$15~\kms.

Assuming that
the molecular gas is in local thermodynamical equilibrium (LTE),
\coo\ column density, \Ncoo, can be computed from the formula 
(see, e.g., Dickman 1978)
 
\begin{equation}
N(^{13}{\rm CO}) =
2.42 \times 10^{14}~\frac{T_{\rm ex}~\int \tau dv}
{1~-~{\rm exp}(-5.29/T_{\rm ex})}~~~~({\rm cm^{-2}}),
\end{equation}
 
\noindent
where \tex\ is the excitation temperature in K,
$\tau$ is the optical depth, $v$ is the velocity in \kms.
The mass of molecular cloud, \mlte, is measured to be
6.20$\times$10$^5$~\msol, 
provided that \coo\ abundance relative to H$_2$ is 2$\times$10$^{-6}$ 
(Dickman 1978).
In this calculation 
we assumed that the cloud boundary is $\int T_R^* dv \simeq 10$ K \kms\
and that \tex\ is 30~K in the central regions of the northern and
southern components and 15~K in the outer regions,
based on our \co\ line data and
the Massachusetts-Stony Brook \co\ Galactic Plane Survey
(Sanders et al. 1986). 
Our mass estimate adopted a mean molecular weight of 2.3 to account for the
contribution of helium.
Table~2 summarizes the physical parameters of the W31 molecular cloud.

\subsection{CS Line results}

Figure 6 exhibits our CS and \css\ line profiles obtained towards 
G10.15$-$0.34 and G10.30$-$0.15. The line parameters of these spectra are
presented in Table 3.
There is a single velocity component along the line of sight and
it is associated with each \uchii\ region.
The center velocity of CS gas is in good agreement with that of \coo\ gas.
The CS J=2$-$1 and J=3$-$2 lines of both \uchii\ regions show 
distinct high-velocity wings. 
This implies that dense gas exists in high-velocity molecular components.
There are a few tens of massive star-forming regions where
high-velocity wings have been observed in the CS lines.
For example, Plume, Jaffe, \& Evans (1992) detected CS J=7$-$6 line 
emission in 104 massive star-forming regions, selected originally by
the presence of an H$_2$O  maser,
and found high-velocity wings in more than 18 out of them.
The full widths of the CS J=2$-$1 lines are about 50~\kms\ at typical rms 
noise level for G10.15$-$0.34 and about 30~\kms\ for G10.30$-$0.15.
These values are similar to those expected from the \coo\ lines, respectively.
The CS J=3$-$2 line of G10.30$-$0.15 has very similar shape and 
full width as the CS J=2$-$1 line, while the CS J=3$-$2 line of  G10.15$-$0.34
has a different asymmetric shape and a smaller (20~\kms) full width 
than the CS J=2$-$1 line. 
The full width of the CS J=2$-$1 line of G10.15$-$0.34
is comparable to that of G5.89$-$0.39, which drives one of the most
energetic outflows in the Galaxy (Harvey \& Forveille 1988).
On the other hand, we can not find high-velocity wings in the CS lines 
observed in the surrounding regions of both \uchii\ regions. 
It seems mainly because the extents of CS outflows 
are $\lesssim$1$'$, taking into account that CO outflows observed in
massive star-forming regions usually have linear sizes of 
$\lesssim$2~pc (Ridge \& Moore 2001; Beuther et al. 2002). 
So high-resolution CS line observations are required 
to confirm whether the high-velocity gas is indeed due to bipolar 
molecular outflows,
and to explore the physical and dynamical properties of the outflows.

Assuming that the excitation temperature and the beam filling factor 
are identical in the CS and \css\ lines in each transition,
we determine the optical depth of \css\ line emission, $\tau_{\rm p}$,
from the peak brightness temperature ratio using the following formula 

\begin{equation}
T_{\rm b}({\rm C^{34}S}) / T_{\rm b}({\rm CS}) = 
\bigl[ 1 - {\rm exp}(-\tau_{\rm p}) \bigr] / 
\bigl[1 - {\rm exp}(-\tau_{\rm p}R) \bigr].
\end{equation}

\noindent
The estimated values are listed in Table 3. Here the CS to \css\
abundance ratio, $R$, was assumed to be equal to 
the terrestrial value (22.5). 

Figure 7 shows integrated \csj\ line intensity maps. 
The CS gas distribution agrees well with the \coo\ gas distribution in
the dense regions of the northern and southern components.
We can see the prominent hollow near G10.30$-$0.15 
in the CS gas distribution as well.
The \uchii\ regions correspond to the peaks of CS cores,
while the compact components without \uchii\ regions do not. 
This seems to be consistent with 
the result of Kim \& Koo (2001) that
the compact components with \uchii\ regions are smaller and denser than
those without \uchii\ regions.
Based on the observation
they suggested that the former are in an earlier evolutionary stage 
than the latter.
We have also found the same trend in several other \uchii\ regions with
extended envelopes that have two or more compact components.
The results will be presented in a separate paper (Kim \& Koo 2002).

If CS line emission originates from regions in LTE and is optically thin,
the CS column density, \Ncs, is given as
 
\begin{equation}
N({\rm CS}) = 1.88 \times 10^{11}~
T_{\rm r}~ {\rm exp}(7.05/{T_{\rm r}})~
\int{{T_{\rm b}} dv}~~~~({\rm cm^{-2}}),
\end{equation}
 
\noindent
where \tr\ is the rotation temperature in K,
\tb\ is the brightness temperature in K,
and $v$ is the velocity in \kms.
If we take a fractional CS abundance to H$_2$ of
1$\times$10$^{-9}$ (Linke \& Goldsmith 1980; Frerking et al. 1980),
the total masses of CS-emitting regions are 2.13$\times$10$^5$ \msol\ for
G10.2$-$0.3 and 0.82$\times$10$^5$ \msol\ for G10.3$-$0.1.
In these calculations
\tr\ was assumed to be 30~K on the basis of the results
of studies on dense cores in massive star-forming regions
(e.g., Snell et al. 1984; Linke \& Goldsmith 1980).
Table 2 lists the physical parameters of the CS-emitting regions.

\subsection{\co\ J=1$-$0 and J=2$-$1 Line Results}

We obtained \co\ J=2$-$1/J=1$-$0 integrated intensity ratio map
for the squared area of $\sim$$5' \times 5'$ centered at G10.30$-$0.15,
after convolving \co\ J=2$-$1 data to the resolution (53$''$)
of \co\ J=1$-$0 data. 
The integrated velocity range was \vlsr~=~0$-$22~\kms.
Figure 8 shows the distribution of \co\ J=2$-$1/J=1$-$0 ratio (grey scale) 
overlaid with the distribution of \co\ J=2$-$1 line integral (contours).
The ratio peaks over the dense cores associated with
the UC and compact \hii\ regions and is low ($\lesssim$1.2) over 
the hollow region.
The estimated values are between 1.0 and 2.0.
The average value is 1.4,
which is much greater than typical value (0.5$-$0.8) 
for molecular clouds in the Galactic disk (e.g., Sakamoto et al. 1997).
Such a high ($>$1.0) ratio is observed towards the areas around 
\hii\ regions in molecular clouds undergoing massive star formation, 
e.g., Orion A molecular cloud (Castets et al. 1990; Sakamoto et al. 1994),
Sgr A molecular cloud (Oka et al. 1996),
and W51B molecular cloud (Koo 1999).

If a molecular cloud has a uniform temperature,
these high ratios can be obtained 
when the molecular gas is warm, dense, and less opaque.
For example, Sakamoto et al. (1994) showed on the basis of 
an one-zone large velocity gradient (LVG) analysis that 
ratios greater than unity are produced when the CO-emitting regions have 
\tex$\gtrsim$40~K, \nhtwo$\gtrsim$1$\times$10$^3$~\cm, 
and $\tau_{21}$$<$5.
In this model the ratio variation is mainly attributed to the density variation
of unresolved clumps that constitute the cloud.
However such a single-temperature model does not seem to be applicable 
to our case,
since the ratio of \coj\ and \cooj\ line intensities, \ratiob,
ranges between 1.3 and 6.7 with an average of 2.7,
indicating that \co\ line emission is optically thick ($\tau_{21}$$\gg$5).
On the other hand,
the high ratios can also be obtained when a dense 
molecular cloud is externally heated by strong far-ultraviolet (FUV) radiation.
Gierens, Stutzki, \& Winnewisser (1992) performed 
a detailed \co\ and \coo\ line radiative transfer analysis
for spherically symmetric clumps irradiated by intense FUV radiation, 
and showed that the ratios $>$ 1 observed in the Orion A molecular cloud 
are easily explained by the temperature and abundance variations within 
the photodissociation region layer of the clumps.
If we adopt their results, we can derive $<$\nhtwo$>$ and $<$\Nhtwo$>$/\delv\ 
of the CO-emitting regions 
from the ratio of \co\ J=2$-$1 and J=1$-$0 line intensities, \ratioa,
and \ratiob\ 
using their Figures 3 and 5.
The \ratioa's are 1.0$-$2.4 with an average of 1.6. 
The average values for \ratiob\ and \ratioa\ suggest that
$<$\Nhtwo$>$/\delv$\gtrsim$1$\times$10$^{22}$ cm$^{-2}$~(\kms)$^{-1}$ and
$<$\nhtwo$>$$\simeq$10$^4$~\cm. 
The \ratioa\ is lower ($\lesssim$1.4) over the hollow region than 
the surrounding regions, while the \ratiob\ is higher ($\gtrsim$4).
This implies, as expected, that $<$\nhtwo$>$ and $<$\Nhtwo$>$/\delv\
are both lower in the hollow region, i.e.,
$<$\Nhtwo$>$/\delv=(0.3$-$1.0)$\times$10$^{22}$ cm$^{-2}$~(\kms)$^{-1}$
and $<$\nhtwo$>$=(1.5$-$6.0)$\times$10$^3$ \cm.

\section{Discussion}

\subsection{Physical Characteristics of the W31 Molecular Cloud}
 
The W31 molecular cloud is classified as a giant molecular cloud (GMC)
based on the physical parameters determined from the \coo\ line observations.
The CS-emitting regions of W31 are much more massive and larger than 
those (10$^2$$-$10$^4$~\msol\ in mass and 0.5$-$5~pc in size)
of molecular clouds associated with Sharpless \hii\ regions,
which are excited by single late O or early B stars
(Zinchenko et al. 1994; Carpenter, Snell, \& Schloerb 1995).
Hence the fraction of dense ($>$10$^4$ \cm) molecular gas may be much higher
in the W31 cloud than in the molecular clouds with Sharpless \hii\ regions. 
For W31 as a whole,
the CS-emitting regions contain 48\% of the total mass and occupy 16\%
of the total area (see Table 2).
In comparison,
these values are, respectively, one order of magnitude greater than 
estimates (2\% and $<$1\%) for the Gemini OB1 cloud complex with
about 10 Sharpless \hii\ regions,
which has a similar total mass to the W31 cloud (Carpenter et al. 1995).
This seems to be consistent with the observation
that more luminous $IRAS$ sources tend to
be associated with more massive CS cores in the Gemini complex, 
suggesting that more massive cores in general form
more massive stars (Carpenter et al. 1995).
Similar results have been found in the Orion B (L1630) molecular cloud 
(Lada 1992) and the Rosette molecular cloud (Phelps \& Lada 1997).
In these molecular clouds the embedded clusters are associated with the
most massive dense cores.
These observations indicate that both high gas density and high gas mass are
required for the formation of stellar clusters.
On the other hand, the CS-emitting regions of W31 are further more massive and 
larger than those of some other GMCs undergoing active massive star formation,
such as M17 (Snell et al. 1984) and Orion B (Lada et al. 1991).
This difference is likely due to difference in evolution
(destruction of dense cores by massive stars)
as well as difference in the initial physical properties, 
since some newly formed massive stars have already emerged as visible objects
in the regions 
(e.g., Stephenson \& Hobbs 1961; Warren \& Hesser 1978).

The ratio of infrared (IR) luminosity to mass is a good measure 
of the massive star formation activity of molecular clouds. 
The ratio varies widely from cloud to cloud, 
and shows little correlation with cloud mass from 10$^2$ to 10$^7$~\msol\
or location in the Galaxy (Mooney \& Solomon 1988; Scoville \& Good 1989;
Carpenter, Snell, \& Schloerb 1990).
We estimate the ratio for the W31 molecular cloud using $IRAS$ 
HIgh RESolution processing (HIRES) images at 60~\mum\ and 100~\mum.
After a smooth background emission is removed from each image,
the total flux densities at 60~\mum\ and 100~\mum, $F_{60}$ and
$F_{100}$, are measured to be
$(7.4\pm0.8) \times 10^4$~Jy and 
$(1.09\pm0.10) \times 10^5$~Jy within the could boundary, 
respectively.
Here we took average intensities on the cloud boundary as the background
emission levels.
The total IR luminosity in the wavelength range 1$-$500~\mum,
\lir, can be calculated from $F_{60}$ and $F_{100}$ in Jy using
the following formula
(Lonsdale et al. 1985; Lee et al. 1996)
 
\begin{equation}
L_{\rm IR} ~= ~0.394 ~R(\overline{T_d}, \beta)
~\bigl[~ F_{100} + 2.58~ F_{60} ~\bigr]~ d^2 ~~~~~(L_\odot),
\end{equation}
 
\noindent
where $T_d$ is the 60/100~\mum\ color temperature in K,
$\beta$ is the index in the emissivity law,
$Q_{\rm abs} \sim \lambda^{- \beta}$,
and $d$ is the source distance in kpc.
The $R(\overline{T_d}, \beta)$ is color correction factor
that accounts for the flux radiated outside the 60~\mum\ and
100~\mum\ $IRAS$ bands. 
If $\beta$ is assumed to be 1 (Hildebrand 1983),
\lir\ is derived to be 5.8$\times$10$^6$~\lsol,
which gives \lir/\mlte=9.4~\lsol/\msol.
This estimate is much greater than 
the average ratio (2.8~\lsol/\msol) of all molecular clouds 
in the inner Galactic plane,
but is comparable to the median value (7~\lsol/\msol) for GMCs with 
\hii\ regions (Scoville \& Good 1989).  

We also derive star formation efficiency,
SFE$\equiv {M_\ast}/{(M_\ast + M_{\rm cloud})}$, of the W31 cloud.
The stellar mass $M_\ast$ was computed  
from the Lynman continuum photon flux \Nc\ measured by radio continuum
observations of G10.2$-$0.3 and G10.3$-$0.1, 
2.7$\times$10$^{50}$ photons~s$^{-1}$ (Kim \& Koo 2001).
Here we adopt
$<M_\ast>/<N'_c> = 5.7 \times 10^{-47}$ \msol~(photons~s$^{-1}$)$^{-1}$,
which was determined by McKee \& Williams (1997) using the initial mass
function of Scalo (1986) and the parameters of OB stars of 
Vacca, Garmany, \& Shull (1996).
The SFE estimated for the entire cloud is 3\%, which is somewhat greater 
than the Galactic median value of $\sim$2\% (Myers et al. 1986).
On the other hand, the SFE's for the two CS-emitting regions 
are derived to be 6\% and 4\%, respectively (Table 3).

\subsection{Interaction between Ionized and Molecular Gas}

Figure 9 shows the ionized gas distribution (contours) superposed on 
the molecular gas distribution (color). For both G10.2$-$0.3 and 
G10.3$-$0.1, the morphology of extended envelope seems to be largely 
determined by the distribution of the ambient molecular gas.
In G10.2$-$0.3, the radio continuum surface brightness decreases slowly
to the east while it drops rapidly to the west. This surface brightness
behavior can be naturally understood if the central part of G10.2$-$0.3 
is situated within the molecular cloud and the eastern protuberance is 
a result of the champagne flow (Israel 1978; Tenorio-Tagle 1979). 
A large ($>$10~\kms) velocity gradient 
exists between the central and eastern parts (see Fig. 2), 
which is compatible with what would be expected from the champagne flow model.
On the other hand, the extended envelope of G10.3$-$0.1 has a bipolar
morphology. Such a morphology can be created 
by the champagne flow in case where the HII region forms in a flat 
molecular cloud (Bodenheimer, Tenorio-Tagle, \& Yorke 1979;
Tenorio-Tagle, Yorke, Bodenheimer 1979).
Indeed the distribution of molecular gas in the region is 
severely elongated in the direction orthogonal to 
the bipolar axis of G10.3$-$0.1 (see also Fig. 7$b$). 
Therefore, the extended envelope of G10.3$-$0.1 is also likely to be
a consequence of the champagne flow.
However, we have not observed any significant velocity difference
between the central region and the envelope.
This might be due to a small inclination.
The velocity of a champagne flow is determined by the pressure contrast at
the cloud boundary and can increase up to about 30~\kms\
(Tenorio-Tagle 1979). 
Since the farthest boundary is $\sim$12~pc (7$'$) from the central region 
for both G10.2$-$0.3 and G10.3$-$0.1,
their dynamical ages may be in the range (4$-$12) $\times$ 10$^5$~yr.
For G10.2$-$0.3, our estimate matches with 
the result of Blum et al. (2001),
who derived a mean age of $\sim$1$\times$10$^6$~yr
for the embedded cluster associated with the eastern compact component 
in the central region, based on their near-IR photometry and spectroscopy. 

A striking feature is the distinctive molecular gas hollow
in the northern flat component of the W31 cloud. 
The hollow is $\sim$4~pc (2.\mins4) in size. It is located close to 
the UC and compact HII regions.
The thickness of the flat molecular cloud is much smaller 
at the hole position than other parts. 
These observations strongly suggest that the hollow might have been produced 
due to the dispersion and destruction of molecular gas 
by the HII regions and their ionizing stars. 
This view is supported by the fact that
the \co\ J=2$-$1/J=1$-$0 integrated intensity ratio exceed unity around the UC
and compact \hii\ regions, implying that the \hii\ regions and their
central stars are closely interacting with the surrounding molecular
materials.
It is well known that the surrounding gas and dust of newly formed
massive stars should be eventually dispersed and destroyed by strong radiation, 
stellar winds, and \hii\ regions.

If the molecular gas column density of the hollow region had been equal
to the average value of the surrounding regions, which have
$\int T_R^* dv \gtrsim 25$ K \kms,
the amount of molecular gas dispersed by the massive stars would be
(4$\pm$1)$\times$10$^3$~\msol. 
The quoted error represents variation in the measured mass
depending on the assumed threshold value for the surrounding regions.
Our estimate is not very different from the ionized gas mass of G10.3$-$0.1,
2$\times$10$^3$~\msol, determined from radio continuum observations 
(Kim \& Koo 2001).
The derived mass and dynamical age imply 
that the disruption rate of the molecular cloud is 
(3$-$10)$\times$10$^{-3}$~\msol~yr$^{-1}$,
which is very similar to the results of two-dimensional
numerical calculations for the champagne flows in various conditions,
(5$-$10)$\times$10$^{-3}$~\msol~yr$^{-1}$
(Yorke, Bodenheimer, \& Tenorio-Tagle 1982).
Whitworth (1979) also performed a semi-analytic study on 
the efficiency with which massive stars destroy their parental molecular clouds
and showed that the disruption rate is determined by 
the Lyman continuum photon flux, the ambient density, and the elapse time.
The Lyman continuum photon flux of G10.3$-$0.1 is 
5.5$\times$10$^{49}$ photons s$^{-1}$ (Kim \& Koo 2001).
The average density of the surrounding regions with 
$\int T_R^* dv \gtrsim 25$ K \kms\ is $\sim$1$\times$10$^3$~\cm.
In that case the disruption rate is
(5$\pm$1)$\times$10$^{-3}$~\msol~yr$^{-1}$.
This value is in good agreement with the rate estimated above, too.
In addition to strong radiation, 
the bipolar outflow associated with G10.30$-$0.15 might have played 
a role in the dispersion process.  
Therefore,
it is most likely 
that the northern flat molecular cloud is being disrupted by 
newly formed massive stars.
This may also be true for the southern spherical molecular cloud,
although compelling observational evidence for molecular gas disruption
have not been found except the champagne flow.

\section{Conclusions}

We have carried out 21~cm radio continuum, \hrrl\ RRL, 
\co, \coo, CS, and \css\ line 
observations of the W31 \hii\ region/molecular cloud complex. 
The physical parameters of the W31 cloud
are largely comparable to those of other GMCs that are 
undergoing active massive star formation.
However, the CS-emitting regions are more massive and larger than
those of some other GMCs with indicators of massive star formation.
The large amount of dense ($>$10$^4$~\cm) gas implies 
that the W31 cloud has ability to form rich stellar clusters
and suggests, together with the young dynamical ages 
($\lesssim$1$\times$10$^6$~yr) of G10.2$-$0.3 and G10.3$-$0.1, 
that star formation has recently begun in the cloud. 
On the other hand, 
we have found high-velocity molecular gas towards G10.15$-$0.34 and 
G10.30$-$0.15, which are likely due to bipolar outflows.
The extended envelopes of both G10.2$-$0.3 and G10.3$-$0.1
may be results of the champagne flows,
based on the distributions of ionized and molecular gas and 
the velocity gradient of RRL emission.
In the vicinity of the UC and compact \hii\ regions in G10.3$-$0.1,
a distinctive molecular gas hollow is present
and the \co\ J=2$-$1/J=1$-$0 line intensity ratio exceeds unity.
Together these observations support that  
the \hii\ regions and their ionizing stars are strongly interacting
with the surrounding molecular materials.
Therefore, W31 is a good example showing not only formation of massive stars
but also destruction of their parental molecular cloud by them.

\acknowledgements
We are grateful to Bob Blum, John Carpenter, Youngung Lee, and 
Young Chol Minh for helpful comments, and Yong-Sun Park for giving us part of
his observing time at the TRAO.
We also thank the staff at the IPAC in Caltech for processing the $IRAS$
HIRES maps for W31 region.
This work has been supported by a bilateral KOSEF-Korea and
CONACyT-M\'exico agreement (2000-113-01-2),
and also BK21 Program, Ministry of Education, Korea through SEES.

\clearpage

\clearpage

\begin{figure}
\vskip 0cm
\figurenum{1}
\epsscale{0.8}
\plotone{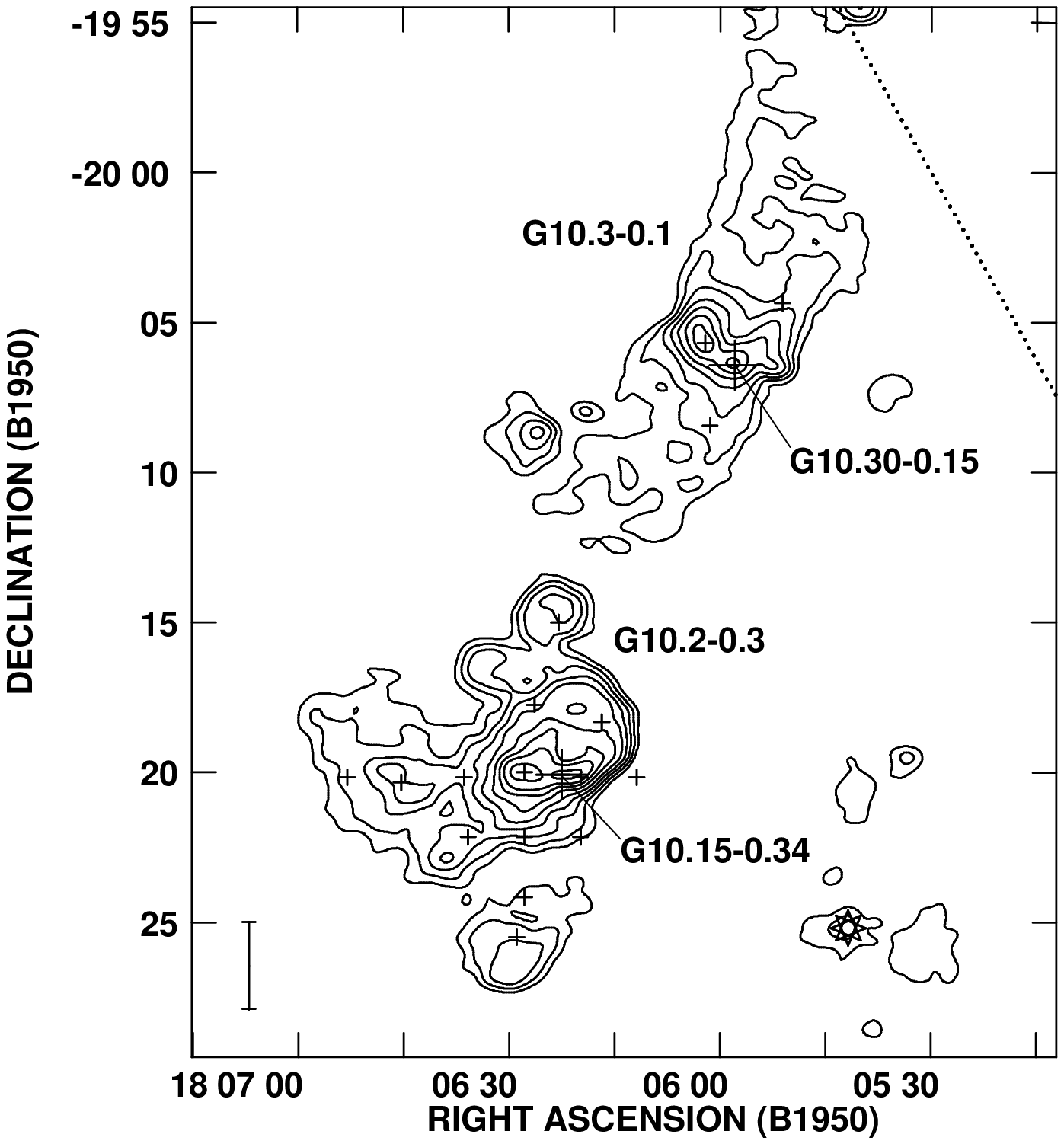}
\vskip -1.0cm
\figcaption[f1.eps]{
21~cm radio continuum image of W31 made with the VLA (DnC array).
This image is corrected for the primary-beam attenuation.
The synthesized beam is $37'' \times 25''$ at P.A.=76$^\circ$.
Contours are 10, 30, 70, 150, 300, 600, 1000, 1500, and 2300~\mjyb.
Large crosses are the positions of G10.15$-$0.34 and G10.30$-$0.15, 
while small crosses are the positions where \hrrl\ line emission was observed.
The SNR G10.0-0.3 is marked by an eight-pointed star 
and the Galactic midplane is represented by the dotted line.
A 5 pc linear scale bar is shown in the bottom left corner.
}
\end{figure}

\clearpage
\begin{figure}
\vskip 0cm
\figurenum{2}
\epsscale{1.0}
\plotone{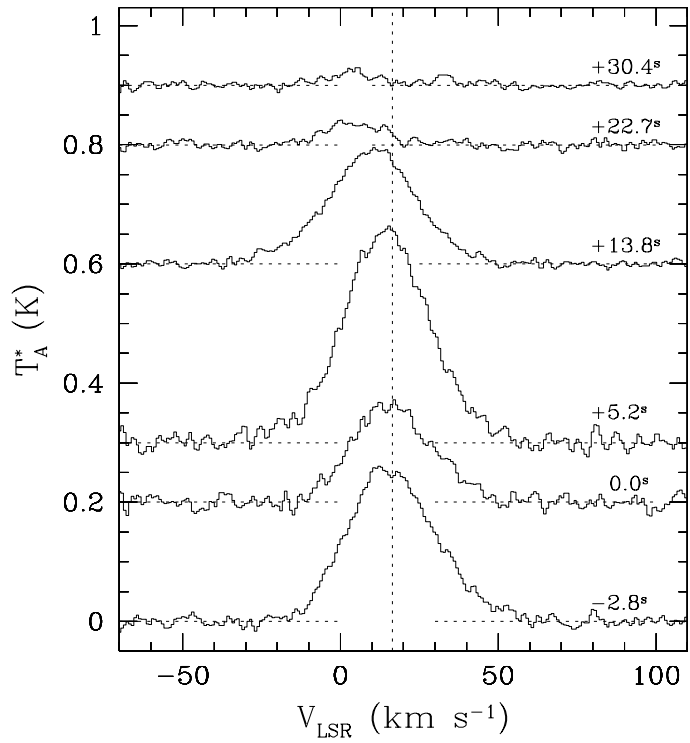}
\vskip 0cm
\figcaption[f2.eps]{
\hrrl\ line spectra observed along
$\delta (1950) \simeq -20^\circ 20' 05''$ in G10.2$-$0.3.
Right ascension offsets with respect to G10.15$-$0.34
are given in second on the right side. The vertical dotted line marks
the center velocity of \hrrl\ line emission observed toward G10.15$-$0.34,
\vlsr=16.4~\kms.
}
\end{figure}

\clearpage
\begin{figure}
\vskip 0cm
\figurenum{3}
\epsscale{1.0}
\plotone{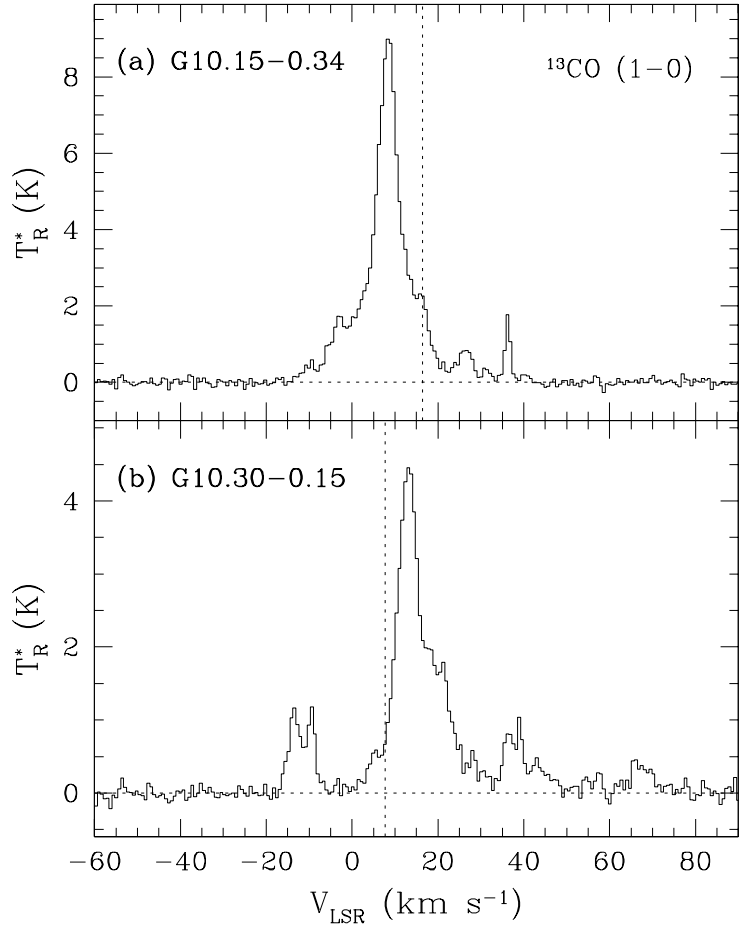}
\vskip 0cm
\figcaption[f3.eps]{
\cooj\ line profiles obtained towards
(a) G10.15$-$0.34 and (b) G10.30$-$0.15.
In each panel, the vertical dotted line indicates the center velocity
of \hrrl\ line.
}
\end{figure}

\clearpage
\begin{figure}
\vskip 0cm
\figurenum{4}
\epsscale{0.8}
\plotone{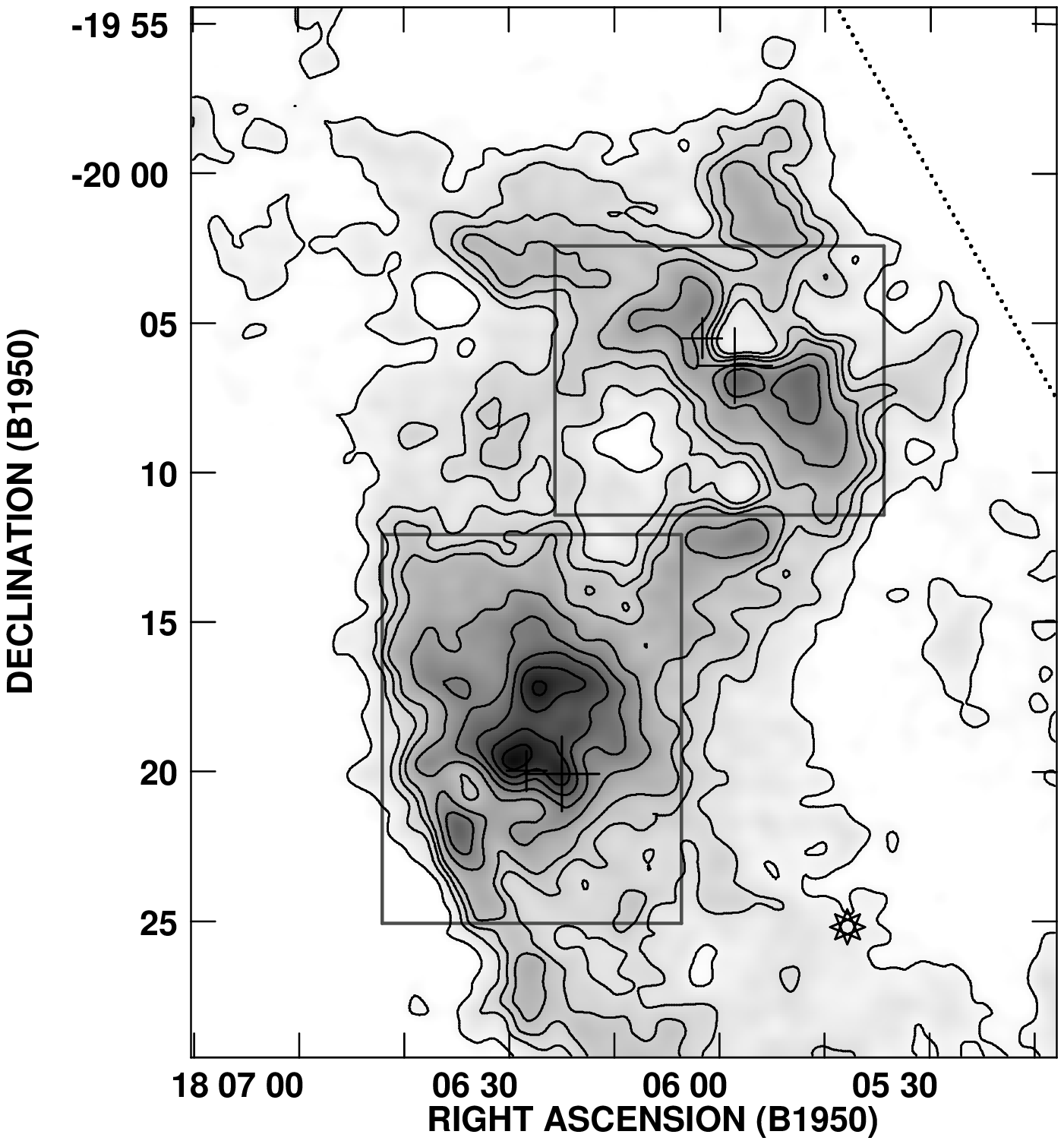}
\vskip -1.0cm
\figcaption[f4.eps]{
\cooj\ integrated intensity map. The integrated velocity range
is from \vlsr=0 to 22~\kms. 
Contour levels are 13, 20, 25, 30, 40, 50,
60, 70 and 80~K~\kms\ and grey scale flux range is 10$-$86~K~\kms.
The boxes show the areas covered by our \csj\ line observations.
Large and small crosses represent the \uchii\ regions and strong
compact components without \uchii\ regions, respectively.
The SNR G10.0-0.3 is marked by an eight-pointed star
and the Galactic midplane is indicated by the dotted line.
}
\end{figure}

\clearpage
\begin{figure}
\vskip 0cm
\figurenum{5}
\epsscale{1.0}
\vskip -1.0cm
\figcaption[f5.eps]{
Channel maps of \cooj\ line emission. The center velocity is given 
in \kms\ at the upper-right corner in each panel. 
Contours are 1, 2, 3, 4, 6, and 8~K and
grey scale flux range is 0.5$-$10.0~K.
Symbols are the same as in Figure 4.
}
\end{figure}

\clearpage
\begin{figure}
\vskip 0cm
\figurenum{6}
\epsscale{1.0}
\plotone{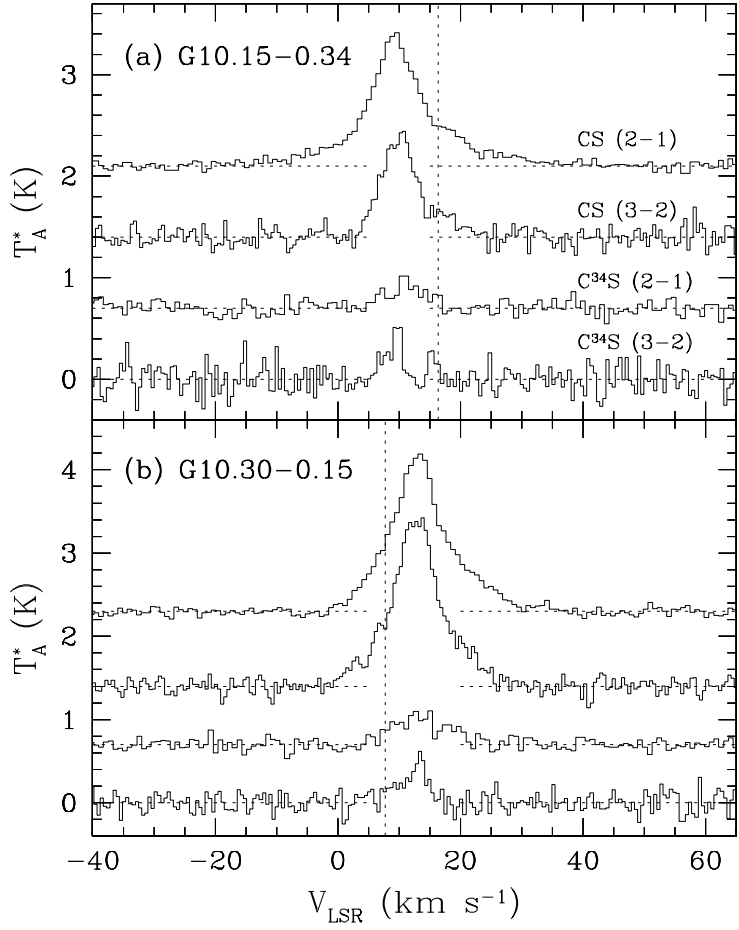}
\vskip 0cm
\figcaption[f6.eps]{
CS J=2$-$1 and J=3$-$2 and \css\ J=2$-$1 and J=3$-$2 line profiles
observed towards (a) G10.15$-$0.34 and (b) G10.30$-$0.15.
The center velocity of \hrrl\ line is marked by the vertical dotted line in
each panel.
}
\end{figure}

\clearpage
\begin{figure}
\vskip 0cm
\figurenum{7}
\epsscale{0.7}
\plotone{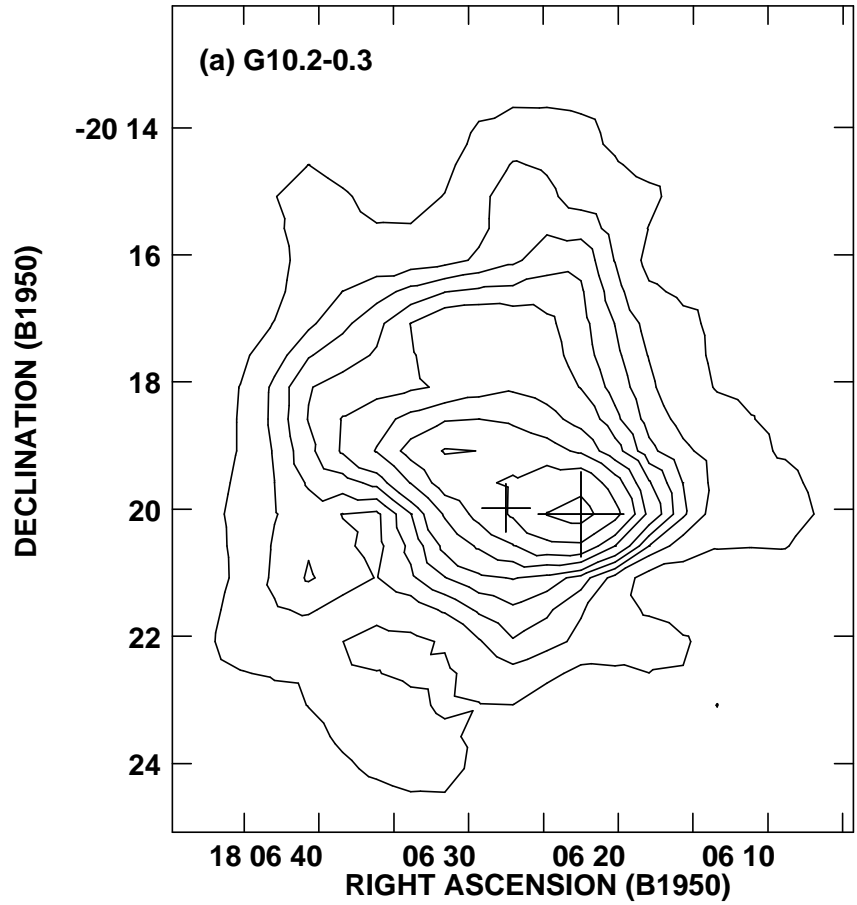}
\vskip 0cm
\figcaption[f7a.eps]{
\csj\ integrated intensity maps of (a) G10.2$-$0.3 and (b) G10.3$-$0.1.
Contour levels are 1.5, 3, 4, 5, 6, 7.5, 9, 11, and 14 K~\kms\ in (a), and
1.2, 2, 3, 4, 7, 12, and 18 K~\kms\ in (b).
Large and small crosses represent the \uchii\ regions and strong compact
components without \uchii\ regions, respectively.
}
\end{figure}

\clearpage
\begin{figure}
\vskip 0cm
\figurenum{7}
\epsscale{0.7}
\plotone{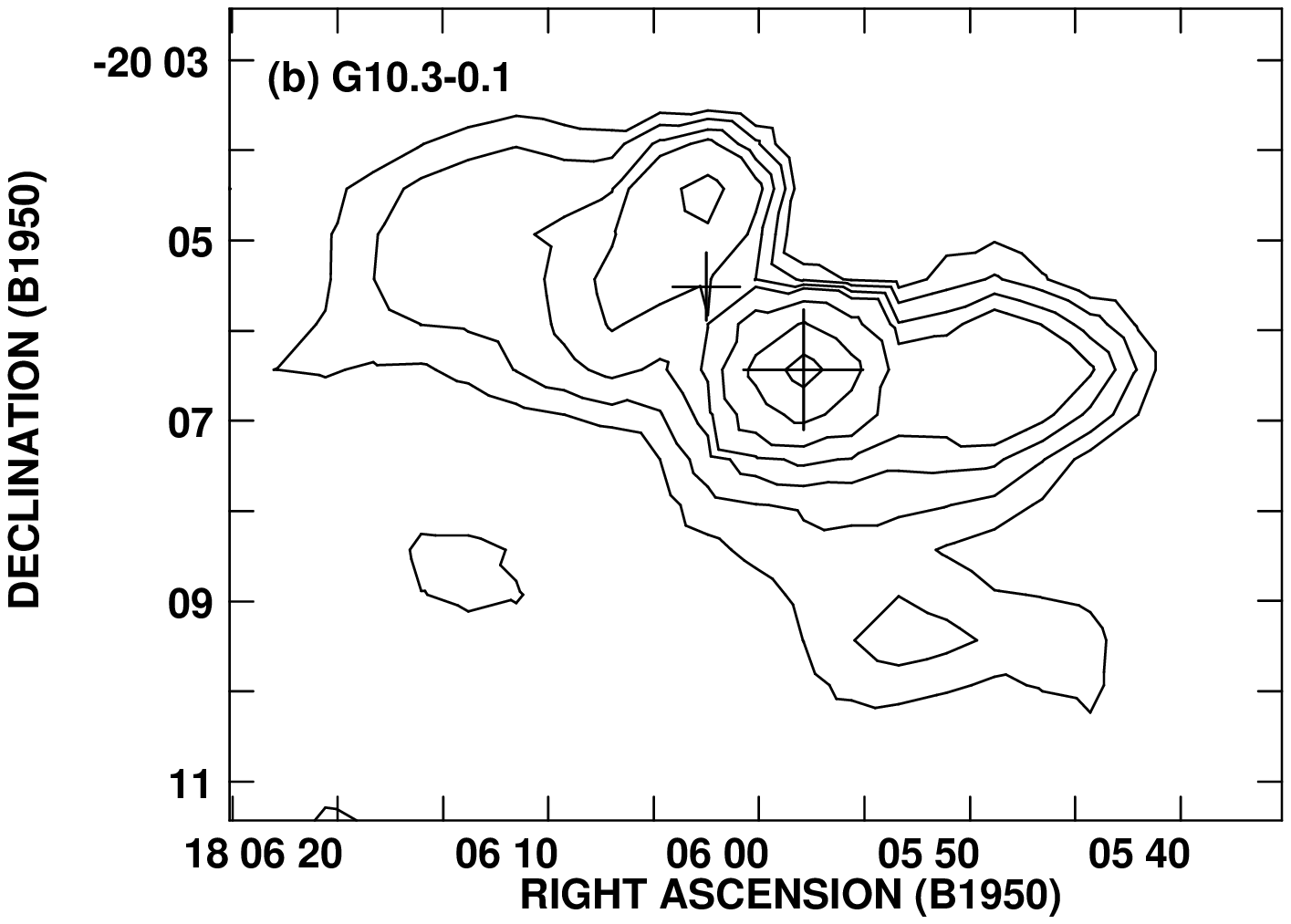}
\end{figure}

\clearpage
\begin{figure}
\vskip 0cm
\figurenum{8}
\epsscale{0.7}
\plotone{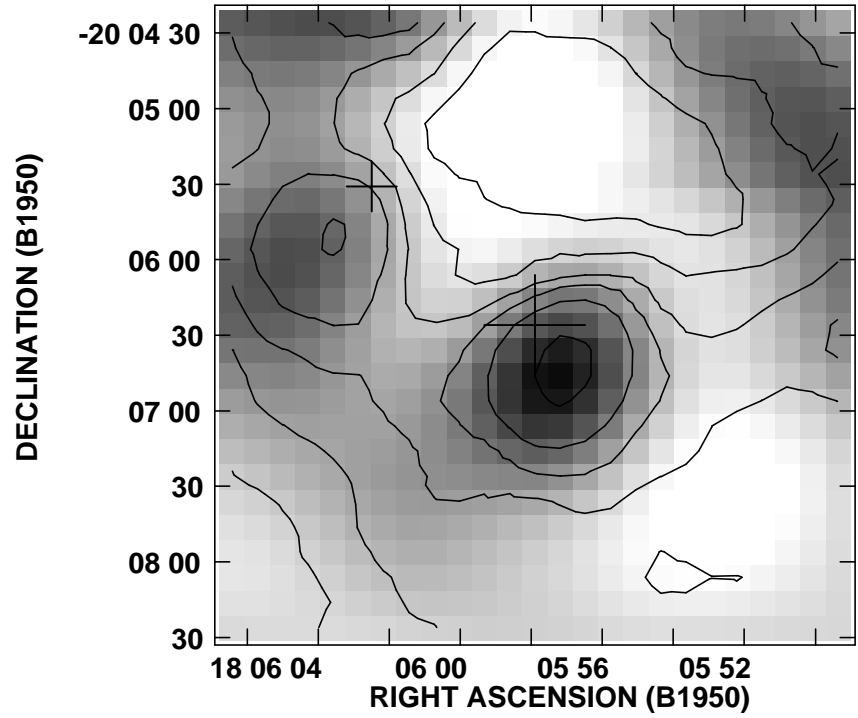}
\vskip 0cm
\figcaption[f8.eps]{
\co\ J=2$-$1/J=1$-$0 integrated intensity ratio map (grey scale) and 
\co\ J=2$-$1 integrated intensity map (contours) of a $\sim$5$'$$\times$5$'$ 
area around G10.30$-$0.15.
Grey scale flux range is 1.2$-$2.0 and 
contours are 80, 100, 120, 140, 170, and 240~K~\kms.
Large cross represents G10.30$-$0.15, while small cross indicates 
the strong compact components without \uchii\ region.
}
\end{figure}

\clearpage
\begin{figure}
\figurenum{9}
\epsscale{0.9}
\vskip 0.0cm
\figcaption[f9.eps]{
\coo\ integrated intensity map (color) and 21~cm radio continuum image 
(contours). The integrated velocity range is 0$-$22~\kms. 
Contour levels are 10, 30, 70, 200, 500, 1000, and 2200~\mjyb.
The color wedge on the right shows the integrated intensity range. 
Crosses represent the positions of G10.15$-$0.30 ($upper$) and
G10.34$-$0.15 ($lower$).
}
\end{figure}

\clearpage
 
\begin{deluxetable}{lrcccccl}
\footnotesize
\tablewidth{0pt}
\tablecaption{MOLECULAR LINES OBSERVED}
\tablehead{
 & \colhead{Frequency} & \multicolumn{2}{c}{Resolution} &
\colhead{FWHM} & & \colhead{OFF} & \colhead{Observation} \\ \cline{3-4}
\colhead{Telescope and Transition} &
\colhead{(GHz)} & \colhead{$\Delta f$ (kHz)} & \colhead{$\Delta v$ (\kms)} & 
\colhead{($''$)} & \colhead{$\eta_{\rm mb}$} & 
\colhead{($l$, $b$)} & \colhead{Date}
}
\startdata
NRAO 12-m & & & & & & \\ 
~~~~~\cooj\ ...... & 110.20135 & ~250 & 
0.68 & 60 & 0.84\tablenotemark{a} &
($10.\!\!^\circ 15$, $-1.\!\!^\circ 50$)\tablenotemark{b} & 1997 June \\
~~~~~\co\ J=2$-$1 ...... & 230.58300 & ~500 & 0.62 & 27 & 
0.51\tablenotemark{a} & 
(10.05, +0.30)\tablenotemark{b} & 2000 March\\
TRAO 14-m & & & & & & \\
~~~~~CS J=2$-$1 .......... & ~97.98095 & ~250 & 0.77 & 60 & 0.48 & 
(10.15, +0.50) & 1997 March\\
~~~~~CS J=3$-$2 .......... & 146.96904 & ~250 & 0.51 & 46 & 0.39 & 
(10.15, +0.50) & 2000 March\\
~~~~~C$^{34}$S J=2$-$1 ....... & ~96.41298 & ~250 & 0.77 & 60 & 0.48 & 
(10.15, +0.50) & 2000 March\\
~~~~~C$^{34}$S J=3$-$2 ....... & 144.61711 & ~250 & 0.51 & 46 & 0.39 & 
(10.15, +0.50) & 2000 March\\
~~~~~\coj\ ...... & 115.27120 & 1000 & 2.60 & 53 & 0.46 & 
(10.05, +0.30)\tablenotemark{b} & 2000 March 
\enddata
\tablenotetext{a}{Corrected main beam efficiency $\eta_{\rm mb}^*$,
$T_{\rm b}$=$T_{\rm R}^*$/$\eta_{\rm mb}^*$.}
\tablenotetext{b}{Reference positions with some \co\ and/or \coo\ emission.}
\end{deluxetable}

\clearpage
 
\begin{deluxetable}{lcccccccc}
\footnotesize
\tablewidth{0pt}
\tablecaption{\cooj\ AND \csj\ LINE OBSERVATIONAL RESULTS}
\tablehead{
  & & \colhead{$R$} & \colhead{$M_{\rm LTE}$} &
\colhead{$n({\rm H_2})$} & \colhead{\delv\tablenotemark{a}} &
\colhead{log~\Nc\tablenotemark{b}} & \colhead{$M_\ast$} & \colhead{SFE}
\\
\colhead{Transition} & \colhead{Source} & \colhead{(pc)} &
\colhead{($10^4$~\msol)} & \colhead{($10^2$~\cm)} & \colhead{(\kms)} &
\colhead{(s$^{-1}$)} & \colhead{($10^4$~\msol)} & \colhead{(\%)}
}
\startdata
\cooj\ ......   & W31         & 24.1 & 62.0 & ~2.0 & ~8.2& 50.43 & 1.54 & 3\\
\csj\ ..........& G10.2$-$0.3 & ~8.0 & 21.3 & 19.2 & 10.0& 50.33 & 1.22 & 6\\
                & G10.3$-$0.1 & ~5.6 & ~8.2 & 21.7 & ~5.4& 49.74 & 0.31 & 4
\enddata
\tablenotetext{a}{FWHM of the average spectrum of the entire mapping area.}
\tablenotetext{b}{Taken from Kim \& Koo 2001.}
\end{deluxetable}

\clearpage

\begin{deluxetable}{lcccc}
\footnotesize
\tablewidth{0pt}
\tablecaption{CS AND \css\ LINE PARAMETERS}
\tablehead{
  & \colhead{\vlsr} & \colhead{\ta} & \colhead{\delv} & $\tau_{\rm p}$
\\
\colhead{Source and Transition} &
\colhead{(\kms)} & \colhead{(K)} & \colhead{(\kms)} & 
}
\startdata
G10.15$-$0.34  & & & & \\
~~~~~CS J=2$-$1 ......... & ~9.6 & 1.31 & 8.8 & \nodata \\
~~~~~CS J=3$-$2 ......... & 10.8 & 1.05 & 7.1 & \nodata \\
~~~~~\css\ J=2$-$1 ...... & 10.4 & 0.32 & 3.1 & 0.28 \\
~~~~~\css\ J=3$-$2 ...... & 10.3 & 0.50 & 2.6 & 0.65 \\
G10.30$-$0.15 & & & & \\
~~~~~CS J=2$-$1 ......... & 13.5 & 1.90 & 8.8 & \nodata \\
~~~~~CS J=3$-$2 ......... & 13.8 & 2.00 & 7.7 & \nodata \\  
~~~~~\css\ J=2$-$1 ...... & 15.1 & 0.40 & 7.0 & 0.24 \\
~~~~~\css\ J=3$-$2 ...... & 13.4 & 0.62 & 2.1 & 0.37 
\enddata
\end{deluxetable}

\clearpage

\end{document}